\documentclass[11pt,twoside,usgraphicx,usenatbib]{article}

\usepackage{asp2006}
\usepackage{psfig}
\usepackage{epsf}
\usepackage{graphics}
\usepackage{lscape}
\usepackage{amssymb}
\usepackage{amsmath}
\usepackage{color}

\newcommand{\be}{\begin{equation}}
\newcommand{\ee}{\end{equation}}

\begin{document}
\title{Feeding the AGN with backflows}   
\author{V. Antonuccio-Delogu\altaffilmark{1,2,3}, J. Silk\altaffilmark{2}}   
\affil{$^{1}$INAF - Osservatorio Astrofisico di Catania, Via S. Sofia 78, I-95123 Catania, ITALY}
\affil{$^{2}$Astrophysics, Department of Physics, University of Oxford, OX3 RH1 Oxford, UK}
\affil{$^{3}$Scuola Superiore di Catania, Via San Nullo, 5/i, I-95123 Catania, Italy}
\begin{abstract}
We study the internal circulation within the cocoon carved out by a relativistic jet emanating from an AGN, first developing model and then validating it using a series of numerical simulations. We notice that a significant increase of \emph{circulation} in this flow arises because gradients in the density and entropy develop near the hot spot, as a consequence of Crocco's vorticity theorem. We find simple solutions for the streamlines, and we use them to predict the mass inflow rates towards the central regions. The 2D simulations we perform span a rather wide range of mechanical jet's input power and Black Hole masses, and we show that the predicted nuclear mass inflows are in good agreement with the theoretical model.\\
\noindent
We thus suggest that these backflows could (at least partially) feed the AGN, and provide a self-regulatory mechanism of AGN activity, that is not directly controlled by, but possibly controls, the star formation rate within the central circumnuclear disk.
\end{abstract}


\section{Introduction}
The presence of compact objects (hereafter \emph{Black Holes}, BHs) in the central dense regions og galaxies is often related to the AGN phenomenon. Moreover, since the original theoretical suggestion by \citet{1998A&A...331L...1S}, significant observational evidence has been accumulated concerning the impact of nuclear activity on the global stellar evolution within the host galaxy \citep[see e.g.][for some recent work]{2006Natur.442..888S, 2007RMxAC..28..109Y, 2007MNRAS.382..960K, 2008MNRAS.388...67K}. In addition to its effect on \emph{global} star formation, the presence of an AGN seems also to be connected with \emph{circumnuclear starbursts} on small (from -parsec to kilo-parsec) scales \citep{2001ApJ...559..147S, 2005ASSL..329..263G, 2007MNRAS.380..949S, 2007ApJ...671.1388D}. However, despite all this observational evidence, the connection between AGN activity, negative/positive feedback on star formation, and local starbursts is not well understood.\\
\noindent
If the main physical agent for this connection is the interaction between the jet and the host galaxy's interstellar medium (hereafter ISM), then it is important to model the physical consequences of the propagation of AGN relativistic jets into the ISM, and the mechanism which feeds the parsec-scale accretion disc around the central BHs. Recent simulations \citep{2007ApJS..173...37S, 2008MNRAS.389.1750A} have begun to  self-consistently model  the feedback of a relativistic jet on its host galaxy, and the global consequences for e.g. blue-to-red cloud migration and downsizing \citep{2009MNRAS.396...61T}. Here we will focus our attention on the \emph{large scale structure of the flow} within the cocoon, and on its consequences for the accretion onto the central BH.

\section[]{Formation of the backflow}
We distinguish 3 different sections in the circulation (see Fig.~\ref{figm1}): propagation near the jet, then from the hotspot along the inner part of the bow shock, and along the meridional plane.
\subsection{Model}
\begin{figure}[!ht]
\plotone{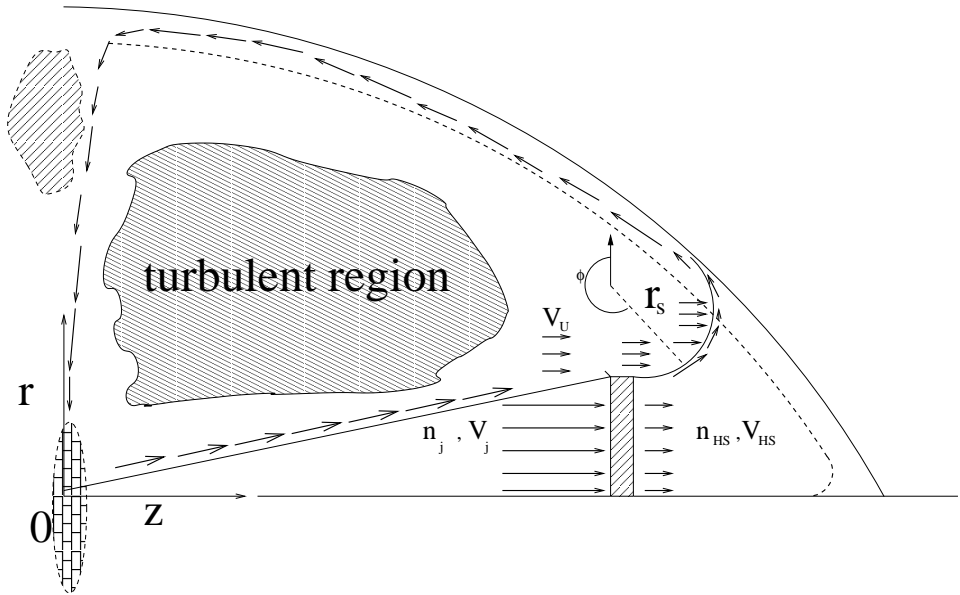}
\caption{Schematic model of the origin of circulation near the hot spot
and in the equatorial plane of the cocoon. We assume that the
\emph{hotspot} (HS) is bounded by a lateral spherical section having
curvature radius $r_{s}$.}
\label{figm1}
\end{figure}

 \citet{1991MNRAS.250..581F} and \citet{1997MNRAS.286..215K} have shown 
 that a \emph{recollimation shock} (hereafter RS) forms at
 some distance along the path of the jet. The post-shocked gas then accumulates into a region confined between the RS and the outer surface of the bow shock (see Figure~\ref{figm1}), with a density $n_{hs}
 \approx 7 n_{j}$, as appropriate to shocked, hot (relativistic) gas. We generically call this region the \emph{hot spot} (HS).\\
\noindent
Most of the cocoon is occupied by 
very low density, high temperature gas, which keeps the jet pressure
confined and collimated, even in the absence of magnetic
fields. In this region the gas is in a turbulent state, with very
little or no systematic motions. This region however does not extend
directly down to the boundary of the jet: the initial shear of the latter induces  parallel motions
of the gas near and outside the jet axis. This gas
eventually reaches the HS, post-shocked region, where the density is significantly
higher than in the cocoon.\\
\noindent
Obviously, the shocked gas within the hot spot has a higher entropy
than that in the cocoon. Under these conditions, \emph{Crocco's theorem} states that  a circulation arises  within a compressible fluid, even if in laminar motion
 \citep[see][for a more recent discussion]{2001...2100..25stzmatost}. In planar geometry the only component of the circulation is normal to the flow plane, and its magnitude is given by:
\be
\omega v = T\frac{ds}{dn} - \frac{dh_{0}}{dn}
\label{eq:bckflw:1}
\ee
where: $n$ is the normal to the direction perpendicular to the streamline, $\omega\equiv\mid\boldmath{\omega}\mid$ is the modulus of the circulation, and $s$ and $h_{0} = h + v^{2}/2$ are the specific entropy and stagnation enthalpy, respectively.\\
\noindent
\subsection{Flow near the Hot Spot}
We will approximate the recollimation shock surface before the hotspot as planar, thus the entropy is constant across the streamline, and the flow will not gain any circulation. However, the gas flowing
\emph{near and outside} the jet, will also eventually reach the HS: the boundary layer separating this region from the turbulent region will be a curved surface, joining the recollimation shock to the bow shock (see Fig.~\ref{figm1}). We then assume that this off-axis flow motion is predominantly directed along the $z-$
direction, thus: $\mathbf{v} \equiv v_{u\mid z}\hat{\mathbf{z}}$. Hereafter, subscripts $u$ and $d$ will denote
quantities computed in the upstream and downstream regions,
respectively, where by the latter we mean the HS, as shown in
Figure~\ref{figm1}. In order to compute $\omega_{hs}$, the
circulation in the hot spot, we will make use of the expression for
the circulation near a curved shock, derived by \citet[][eq. 8]{1958ZAMP...9..637}:
\be
\omega_{hs} = \frac{2v_{z\mid u}}{r_{s}\left(\gamma + 
  1\right)}\frac{\left(M_{n}^{2}-1\right)\mid\cos\phi\mid}{M_{n}^{2}[2+\left(\gamma-1\right)M_{n}^{2}]}
\label{eq:bckflw:2}
\ee
 where: $M_{n} = (v_{z}/c_{s})\cos\theta$ is the Mach number in the
 upstream flow region. As we show in Antonuccio-Delogu and Silk (2009, submitted) one can obtain an exact solution in 2D spherical coordinates for the velocity field after the shocked layer:
\be
v_{\phi}(r_{s}) = \frac{4v_{z\mid u}}{\left(\gamma + 
  1\right)}\frac{\left(M_{n}^{2}-1\right)\mid\cos\phi\mid}{M_{n}^{2}[2+\left(\gamma-1\right)M_{n}^{2}]}
\label{eq:bckflw:3}
\ee 
where $\phi$ is a polar angle and. We see that the azimuthal component of velocity after the shock becomes negative, and the polar component in the downstream
region acquires a positive value: thus, a \emph{backflow} develops.
\begin{figure}[!ht]
\plottwo{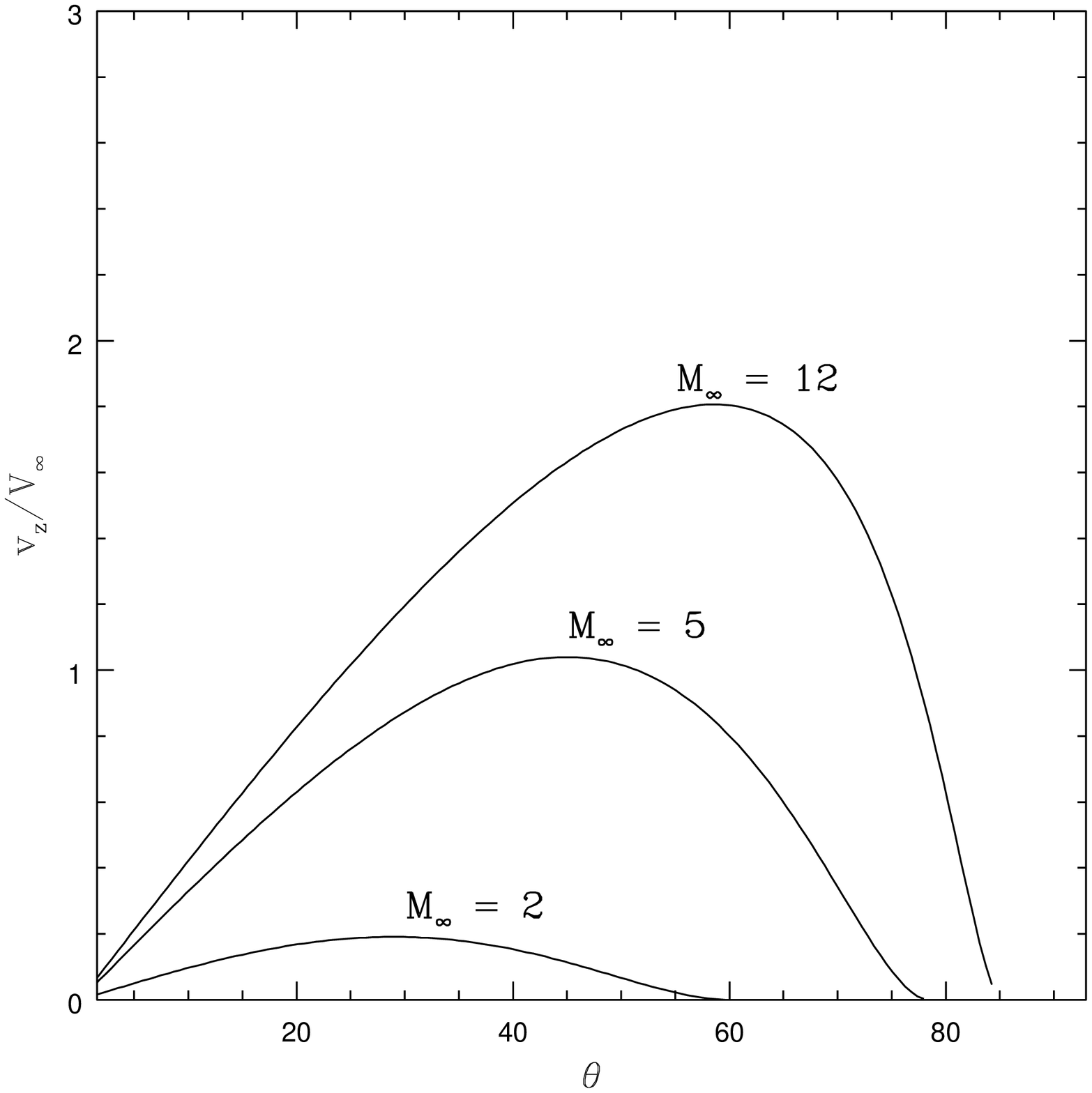}{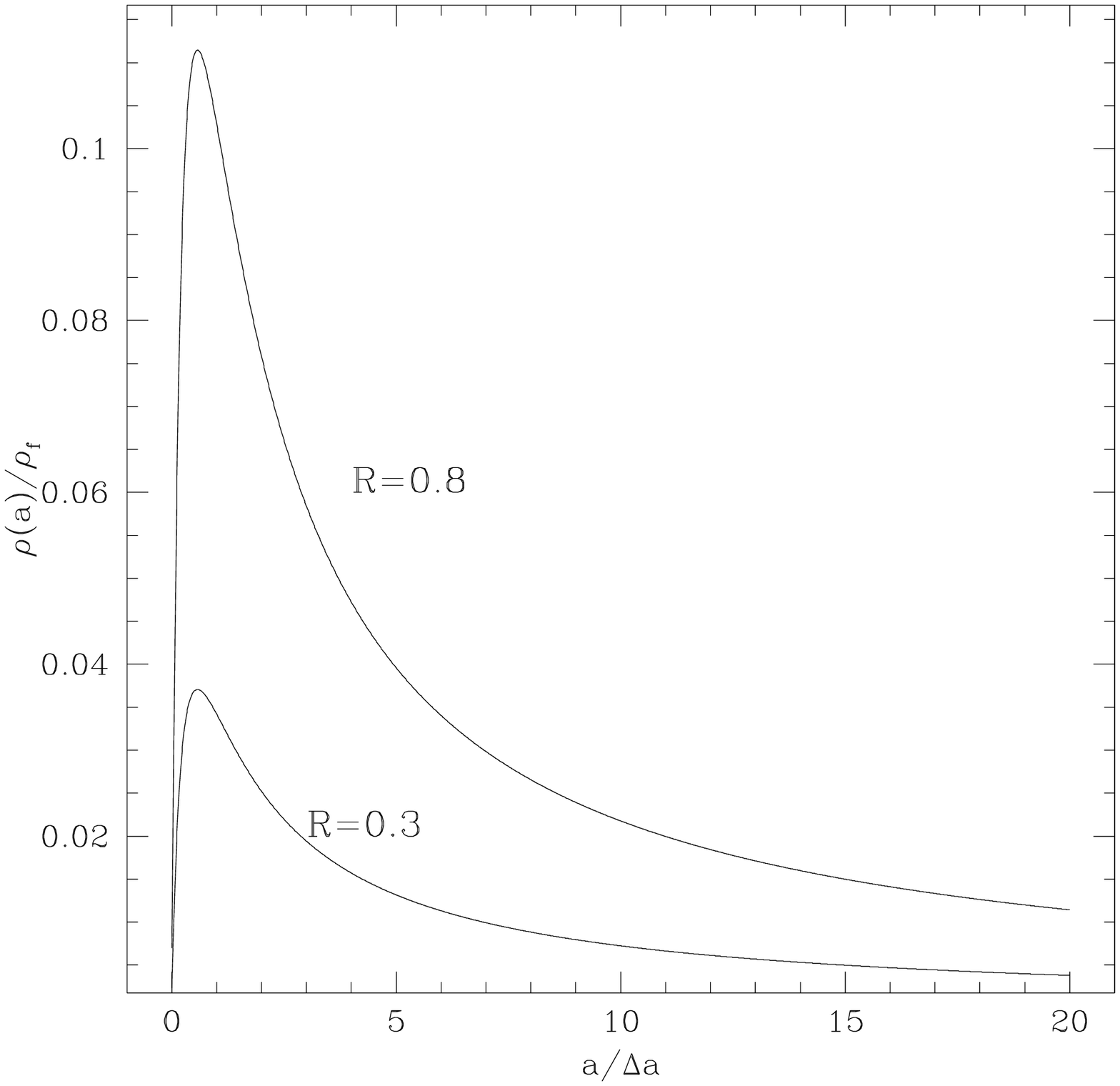}
\caption{\emph{Left:} The excursion of $\mid v_{d\mid z}/v_{u\mid z}\mid$, as
  measured in eq.~\ref{eq:bckflw:3}, for different values of the
  upstream Mach number. The flows within the cocoon usually are
  transonic, with $M_{\infty} \simeq 1-2$. \emph{Right:}The scaled density within the bow shock as a function of
  $a/\Delta a$, the semi-major axis measured in units of the width of
  the spheroidal shell containing the shock. We plot the ratio
  $\rho/(3\rho_{0}a_{0}^{2}/(\Delta a)^{2})$, for two different values
  of the aspect ratio $R$.}
\label{fig_v_theta}
\end{figure}

\subsection{The flow along the bow shock}
\begin{figure}[!ht]
\plottwo{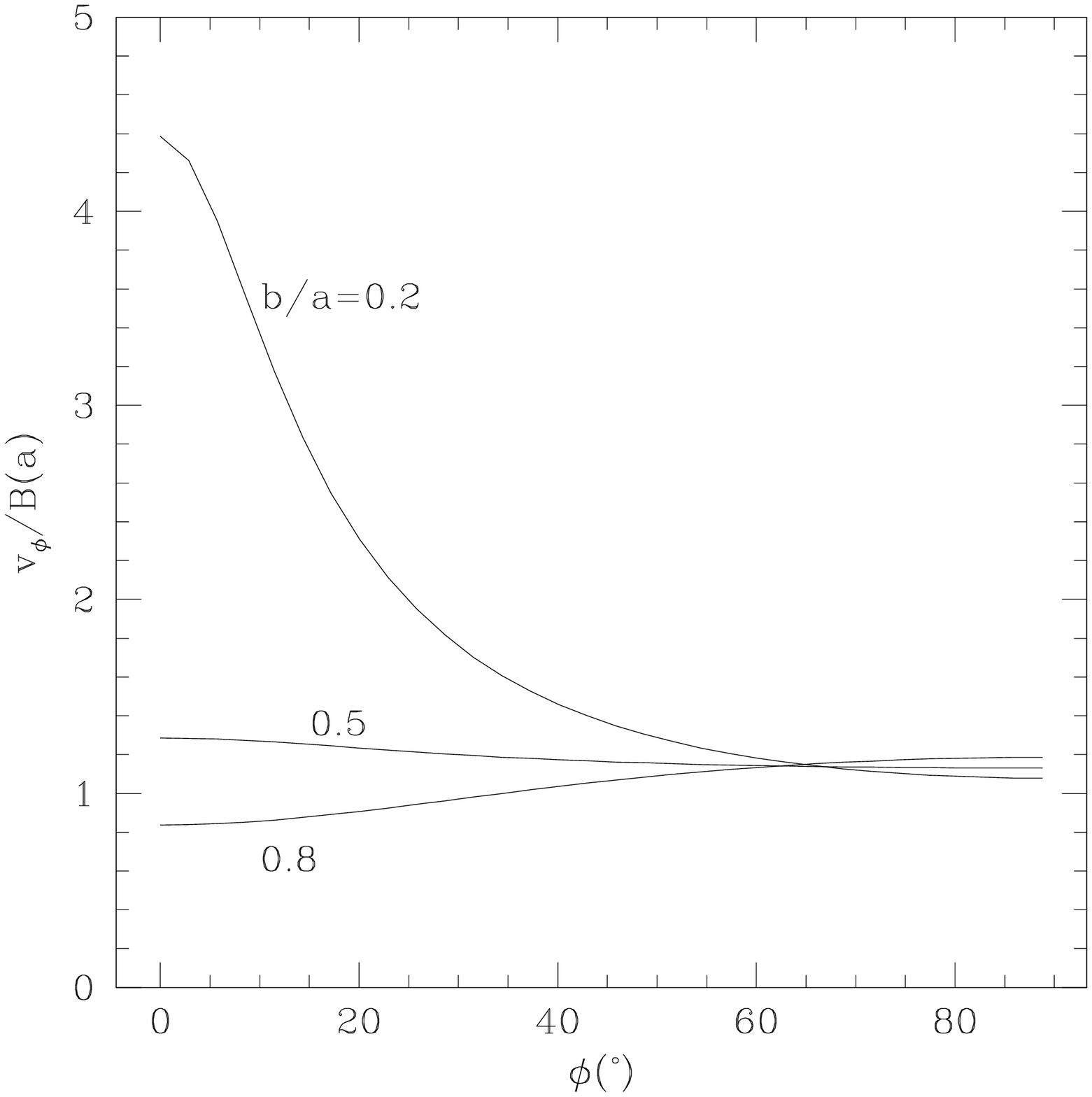}{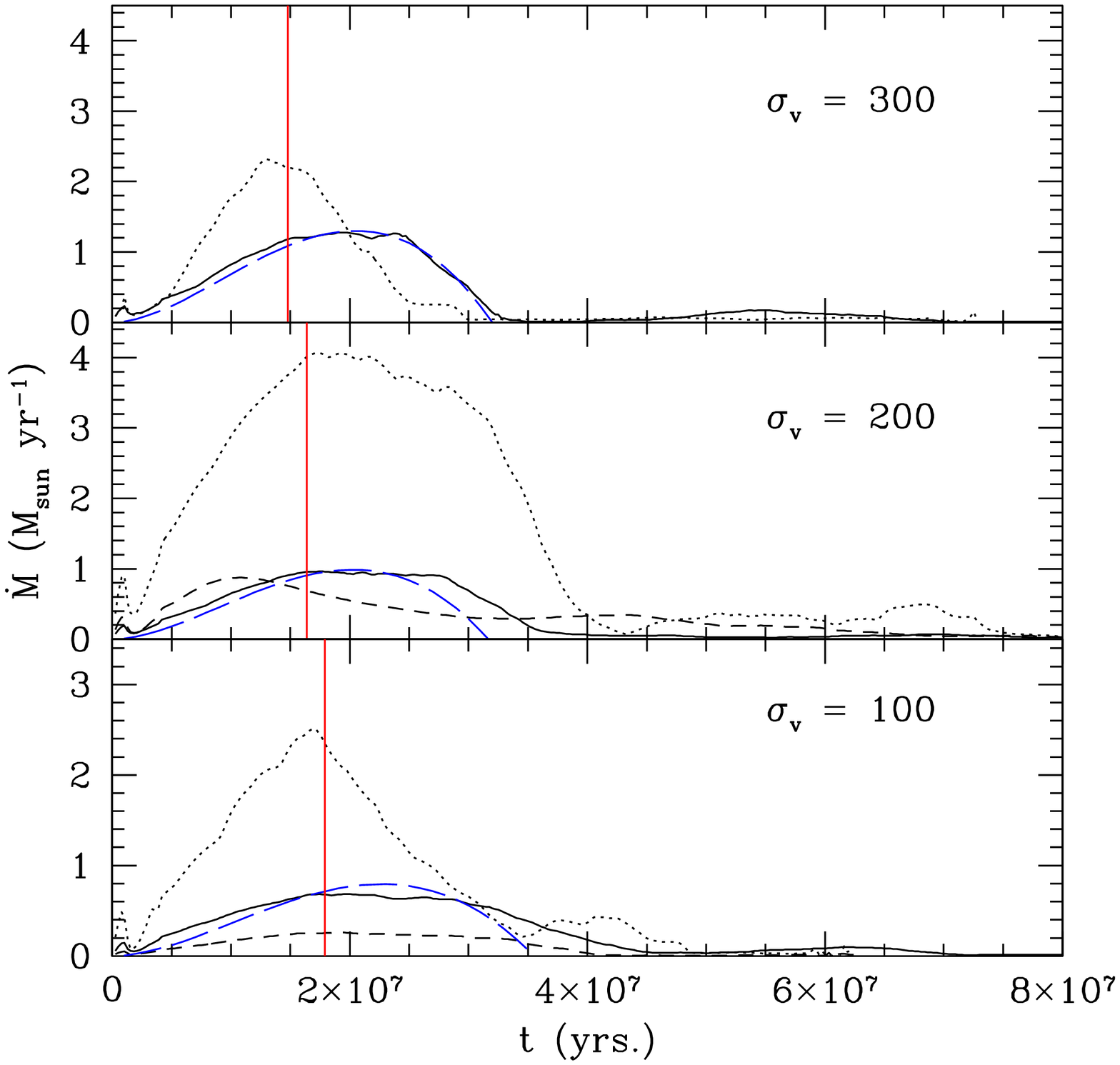}
\caption{(\emph{Left}): Velocity along the bow shock, for different
  aspect ratios.(\emph{Right}): Mass inflow rates around a central disk region. The continous
   lines are for simulations \emph{s(100,200,300)av}, dotted lines for
   \emph{s(100,200,300)p1}, dashed lines for
   \emph{s(100,200,300)m1}. The dashed lines are the fits from
   the model given in the text. The red vertical lines mark the epoch
   when the reconfinement shock is destroyed and the jet starts to
   propagate freely, being only confined by the cocoon's ram pressure.}
\label{fig_4-5}
\end{figure}

After the HS, the backflow enters into the inner part of the bow shock (Figure~\ref{figm1}), and the magnitude of
the rotation is determined by the conservation law for vorticity in 2D: 
\be
\frac{\mid\boldmath{\omega}\mid}{\rho} = const
\label{eq:bckflw:4}
\ee
Thus, in order to determine $\mid\boldmath{\omega}\mid$, we have to determine the density inside the bow shock. Assuming that all the matter which was inside the cocoon is compressed within the bow shock, the final result reads:
\be
\rho_{bs} = 3\rho_{0}\left(\frac{a_{0}}{\Delta
  a}\right)^{2}g(R)\frac{\lambda}{3\lambda^{2} + 3\lambda + 1}
\label{eq:bckflw:8}
\ee
where $\rho_{0}$ is the central density of the galaxy halo, and we have defined: $\lambda = a_{i}/\Delta a$, where $a_{i}, \Delta a$ are the semimajor axis and the width of the bow shock, respectively. Note that the time-dependance
of the density enters through the semi-major axis ($a\equiv a(t)$). Integrating eq.~\ref{eq:bckflw:8} we obtain the velocity along the bow shock, $v_{\phi}$: the result is shown in Fig.~\ref{fig_4-5} (\emph{left}). We again notice that the
maximum variations are always around unity, along the streamlines, and the velocity tends to decrease or to stay almost constant, as we approach the polar regions.\\
\noindent
In summary, our model makes a few predictions. First, at least during
the initial phases of the expansion, a circulation arises inside the
cocoon, induced by the presence of a high density and entropy region (the Hot Spot). The backflow which develops tends to follow the bow
shock, and then bends back near the meridional axis, perpendicular
to the jet. For symmetry reasons, this flow will converge back towards
the jet.

\subsection{Numerical simulations}
The numerical simulations confirm the numerical model sketched (Fig.~\ref{fig_4-5}, \emph{right}). 
\noindent
\begin{figure}[!ht]
\plottwo{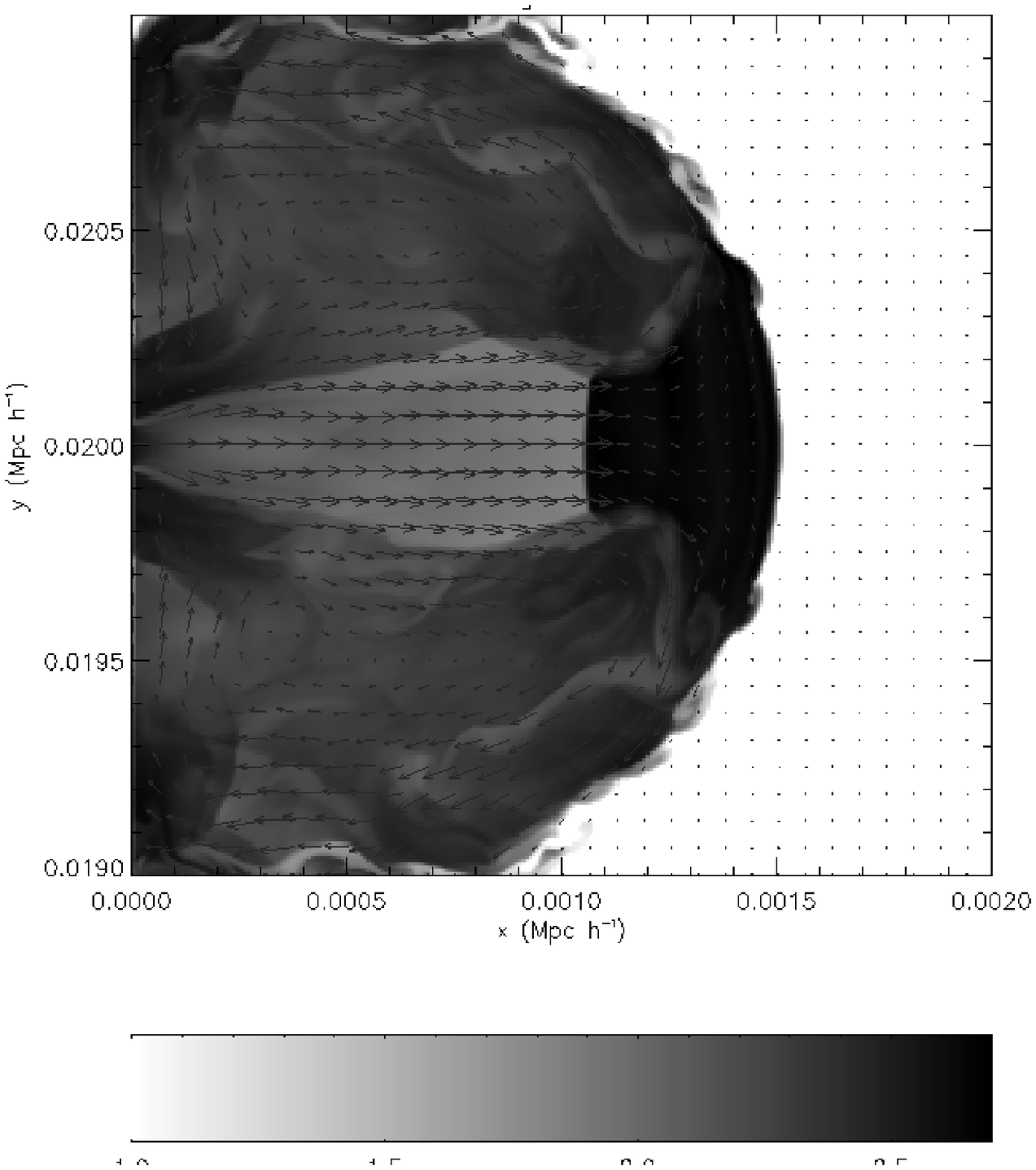}{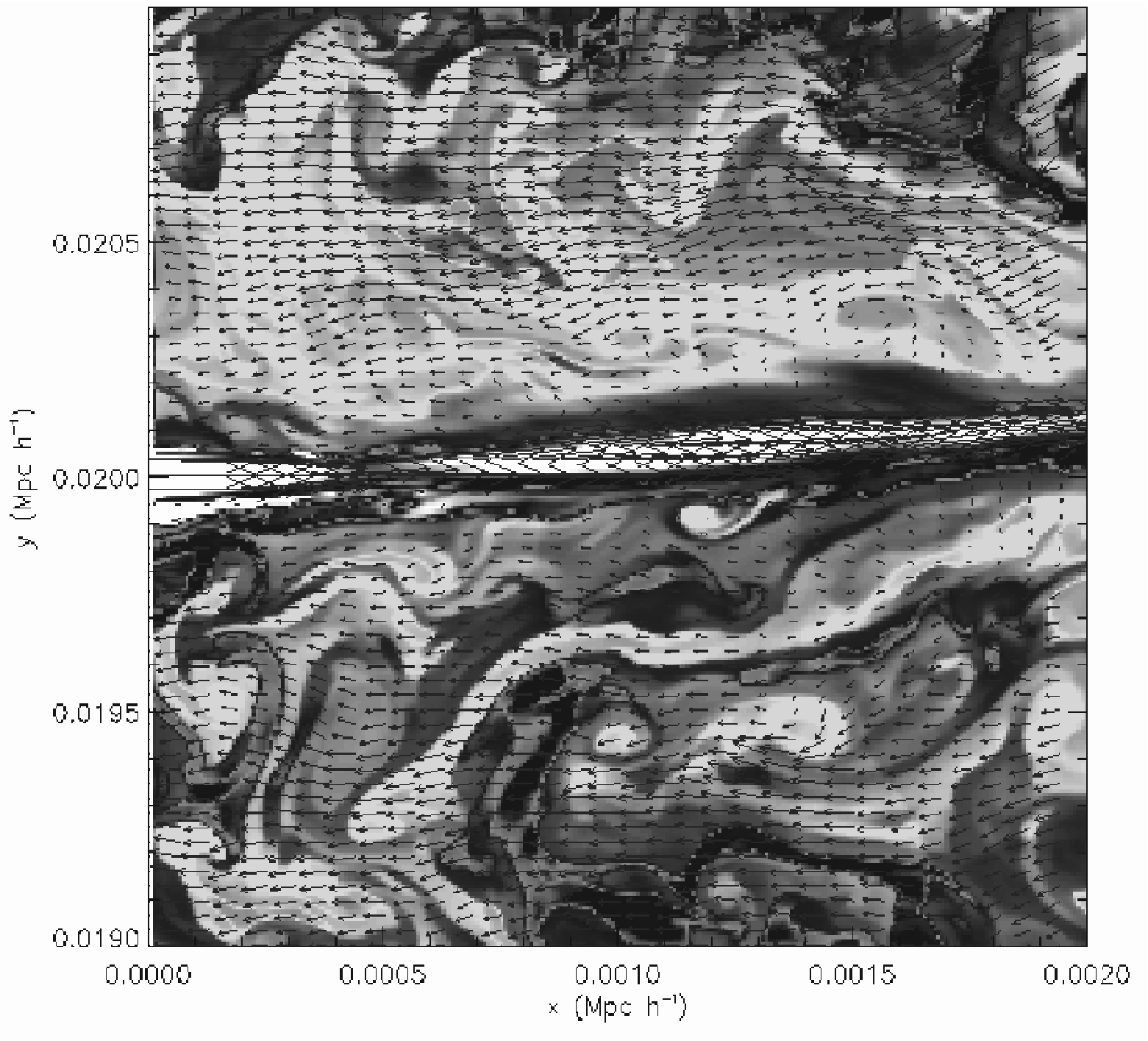}
\caption{(\emph{Left}): Velocity field before the destruction of the jet's shockby KH instabilities. The general pattern is well described by the model sketched in this paper. (\emph{Right}): pattern of the flow after the disappearence of the shock along the jet. Note that the colour scale (not shown) is much smaller than in the figure on the right, thus density differences are in reality very small.}
\label{fig_vsimn}
\end{figure}
It is interesting to observe that the agreement with the model predictions is quite satisfactory, for a wide range of central densities and jet's injection powers. Note that the in these simulations the main parameters like central halo mass, BH mass, jet's injection power, are chosen according to scaling relations (see Antonuccio-Delogu and Silk, 2008, for details).\\
\noindent
The simulations also show that two different circulation regimes describe
 the flow within the cocoon (see Fig.~\ref{fig_vsimn}). The presence of the shock before the HS determines a flow similar to that described in our model. However, after a while this shock is destroyed by Kelvin-Helmholtz instabilities, and the hotspot propagates freely (Fig.~\ref{fig_vsimn}, \emph{right}). The gas near the jet is now reflected by the HS, and determines a counterstreaming flow which propagates back towards the central disc. The net mass inflow rate is however much less.

\section{Conclusion}
Realistic simulations of the flow pattern within an expanding cocoon show that a \emph{backflow} develops soon after the cocoon forms. This determines a coherent flow towards the central parts of the AGN, whose magnitude is of the order of few $M_{\odot}\, \rm{yr}^{-1}$, i.e. the right order to feed the AGN and sustain its activity. A more detailed investigation of the consequences of this backflow is presented elsewhere (Antonuccio-Delogu and Silk, 2009, submitted).

\acknowledgements 
The work of V.A.-D. has been supported by the European Commission,
under the VI Framework Program for Research \& Development, Action
``{\em Transfer of Knowledge}'' contract MTKD-CT-002995 (``{\em
Cosmology and Computational Astrophysics at
  Catania Astrophysical Observatory}''). V.A.-D. would also
express his gratitude to the staff of the 
subdepartment of Astrophysics, Department of Physics, University
of Oxford, for the kind hospitality during
the completion of this work.



\bibliographystyle{mn2e}
\bibliography{biblio}

\end{document}